\documentclass[twocolumns,structabstract]{aa}
\usepackage{amsmath}
\usepackage{graphicx}
\usepackage{txfonts}
\usepackage{natbib}
\usepackage{mathtools}
\usepackage{color}
\usepackage{hyperref}

\bibpunct{(}{)}{;}{a}{}{,} 

\hypersetup{draft}
\begin{document}

\title{Electromagnetic induction heating as a driver \\of volcanic activity on massive rocky planets}

   \author{Kristina Kislyakova
          \inst{1}
          \and
          Lena Noack
          \inst{2}
          }


   \institute{University of Vienna, Department of Astrophysics,  T\"{u}rkenschansstrasse 17, A-1180 Vienna, Austria
   \and   
   Freie Universit\"at Berlin, Malteserstrasse 74-100, 12249, Berlin, Germany}

   \date{Received \today}

\abstract{}
{We investigate possible driving mechanisms of volcanic activity on rocky super-Earths with masses exceeding 3-4~M$_\oplus$. Due to high gravity and pressures in the mantles of these planets, melting in deep mantle layers can be suppressed, even if the energy release due to tidal heating and radioactive decay is substantial. Here we investigate whether a newly identified heating mechanism, namely induction heating by the star's magnetic field, can drive volcanic activity on these planets due to its unique heating pattern in the very upper part of the mantle. In this region the pressure is not yet high enough to preclude the melt formation.}
{Using the super-Earth HD~3167b as an example, we calculate induction heating in the planet's interiors assuming an electrical conductivity profile typical of a hot rocky planet and a moderate stellar magnetic field typical of an old inactive star. Then we use a mantle convection code (CHIC) to simulate the evolution of volcanic outgassing with time.}
{We show that although in most cases volcanic outgassing on HD~3167b is not very significant in the absence of induction heating, including this heating mechanism changes the picture and leads to a substantial increase in the outgassing from the planet's mantle. This result shows that induction heating combined with a high surface temperature is capable of driving volcanism on massive super-Earths, which has important observational implications.}{}


\titlerunning{Induction heating as a driver of volcanic activity}
\authorrunning{K.~G. Kislyakova \& L. Noack}
\maketitle

\section{Introduction}
\label{intro}

Volcanic outgassing is one of the major processes filling a planet's atmosphere with gases. An atmosphere can also be produced during the planet's accretion phase by rapid outgassing from the interior during the solidification of a magma ocean caused by giant impacts. If the surface temperature is high, a magma ocean can be produced via strong irradiation from the host star, as has been proposed for the planet 55~Cnc~e \citep{Demory16}. If stellar short wavelength radiation and wind lead to rapid loss of the atmosphere to space, the atmosphere has to be replenished by volcanic outgassing (e.g., \citealp{Godolt19}). This is especially relevant for close-in planets with an efficient atmospheric escape. In this paper, we focus on atmosphere formation due to volcanic activity.

Sufficient energy release inside a planet's mantle can melt it and drive very strong volcanism. As an example, Io's tidal surface heat flux of 2~W/m$^2$ makes it the most volcanically active body in the solar system despite its relatively small size \citep{McEwen98}. The presence and composition of a planet's atmosphere play a very important role for the atmospheric characterization. With the upcoming observational facilities such as the Atmospheric Remote-sensing Infrared Exoplanet Large-survey (ARIEL) and the James Webb Space Telescope (JWST), understanding the possible formation mechanisms of atmospheres is of pivotal importance.

Recently, \cite{Noack17} and \cite{Dorn18} have shown that at low masses (below 3-4 M$_\oplus$) outgassing positively correlates with the planet's mass since it is controlled by the mantle's volume. At higher masses, outgassing decreases with the planet's mass, which is due to the increasing pressure gradient that limits melting to shallower depths. Observations of small rocky exoplanets with masses lower than 3-4 M$_\oplus$ are extremely challenging. Although ARIEL will be able to observe some small planets, most of its rocky targets will be more massive than the Earth (e.g., \citealp{Edwards19}). Therefore, many planets in the ARIEL and JWST target lists can be potentially affected by insufficient outgassing, and thus a lack of observable atmosphere. In this paper we investigate whether a newly identified heating mechanism, namely heating by electromagnetic induction recently suggested by \citet{K17}, can change this pattern.

Induction effects arise when a planet is embedded in a varying stellar magnetic field and influence the entire planet, from the interior to the atmosphere, and can have observable effects. Induction heats the planet by alternating currents generated in the planetary mantle and drives strong volcanic activity. For planets orbiting M~dwarfs with magnetic fields of a few hundred gauss or more or for planets orbiting very close to stars with solar-like magnetic fields, induction heating of the interiors can be as powerful an energy source as tidal heating \citep{K18,GK20}. In this paper, we focus on rocky super-Earths orbiting close to their host stars with moderate magnetic fields. This paper is organized as follows: Section~\ref{methods} briefly describes our methods, Section~\ref{results} presents our results, and Section~\ref{conclusions} summarizes our conclusions.


\section{Methods}
\label{methods}

 For induction heating to be substantial, two factors are necessary:   a strong magnetic field at the planet's orbit and  a periodic variation of the amplitude of this field. The period of the field variation is determined by the star's rotation, by the planet's orbital period, or by a combination of the two. The star's rotation leads to induction effects in planets if the star's magnetic dipole axis is inclined with respect to its rotation axis. Observations and magnetic maps show that any angle between the two axes is possible \citep{Lang12,Morin10,Fares13,Fares17}. In this study, we investigate the induction effects in HD~3167b, which is a transiting super-Earth with a mass of 5.7~M$_\oplus$ and a radius of 1.57~R$_\oplus$.  Interesting features of this planet are its polar orbit and its close distance to the host star, which are prerequisites for efficient induction heating and resulting volcanic activity. Parameters of the system are summarized in Table~\ref{t_param}. We assume that the star's magnetic dipole axis coincides with the star's rotation axis. The inclination of the orbit of HD~3167b with respect to the star's rotation axis is unknown; however, \citet{Dalal19} found that the second planet in the system, HD~3167c, is on a nearly polar orbit. Since most planetary systems are coplanar and both planets are transiting, HD~3167b should also have a polar orbit. We assume an orbital inclination of 85$^\circ$. Such high orbital inclination can cause a strong induction heating due to a variation of the star's magnetic field at the planet's location due to the planet's motion around the star \citep{K18}. To calculate the magnetic field at a given orbital distance, we use a potential field source-surface model, which is commonly used to model stellar magnetic fields \citep{JohnstoneThesis}, as  described in \citet{K17,K18}.

\begin{table}
\caption{Parameters of the star HD~3167 and planet HD~3167b adopted from \citet{gandolfi17}.}
  \begin{tabular}{l l }
\hline
Parameter    & Value \\
\hline
Stellar mass (M$_\odot$) & 0.877 $\pm$ 0.024 \\
Stellar radius (R$_\odot$) & 0.835 $\pm$ 0.026 \\
Stellar age (Gyr) & 5.0 $\pm$ 4.0   \\
Stellar equilibrium temperature (K)  & 5286 $\pm$ 40  \\
Stellar rotational period (days) & 23.52 $\pm$ 2.87 \\
Stellar magnetic field\footnote{1} (G) & 1--10 \\
Planetary mass (M$_\oplus$)   & 5.69 $\pm$ 0.44  \\
Planetary radius (R$_\oplus$)   & 1.574 $\pm$ 0.054 \\
Planetary equilibrium temperature (K) & 1759 $\pm$ 20  \\
Semi-major axis (au)  & 0.01752 $\pm$ 0.00063  \\
Orbital inclination\footnote{2} (degree) & 85 \\
\hline
\label{t_param}
\end{tabular}
\end{table}
\addtocounter{footnote}{-1}
\footnotetext{Following the magnetism trends presented by \citet{Vidotto14}.}
\addtocounter{footnote}{1}
\footnotetext{Between the stellar rotational axis and the planet's orbital plane.}

We use the code developed to calculate induction heating in planetary interiors, which has been applied to exoplanets with Earth-like conductivity profiles \citep{K17,K18,GK20}. The code is based on the method presented in \citet{Parkinson83} for a sphere with a non-uniform electrical conductivity. The code divides the planetary mantle into layers and solves the induction equation in every layer to find the magnetic field and current. The energy release within each layer can be determined from the current and conductivity. For a simple stellar dipole field, the field around a planet can be considered uniform due to stretching of the magnetic field lines by the stellar wind. Then the current flowing inside the planet has a simple form of a circular current that is  the strongest near the planet's equator, where the equator is defined by the direction of the stellar magnetic field. The magnitude of the current and the heating decreases both with depth and latitude, as is typical of the skin effect.

\begin{figure}
\includegraphics[width=1.0\columnwidth]{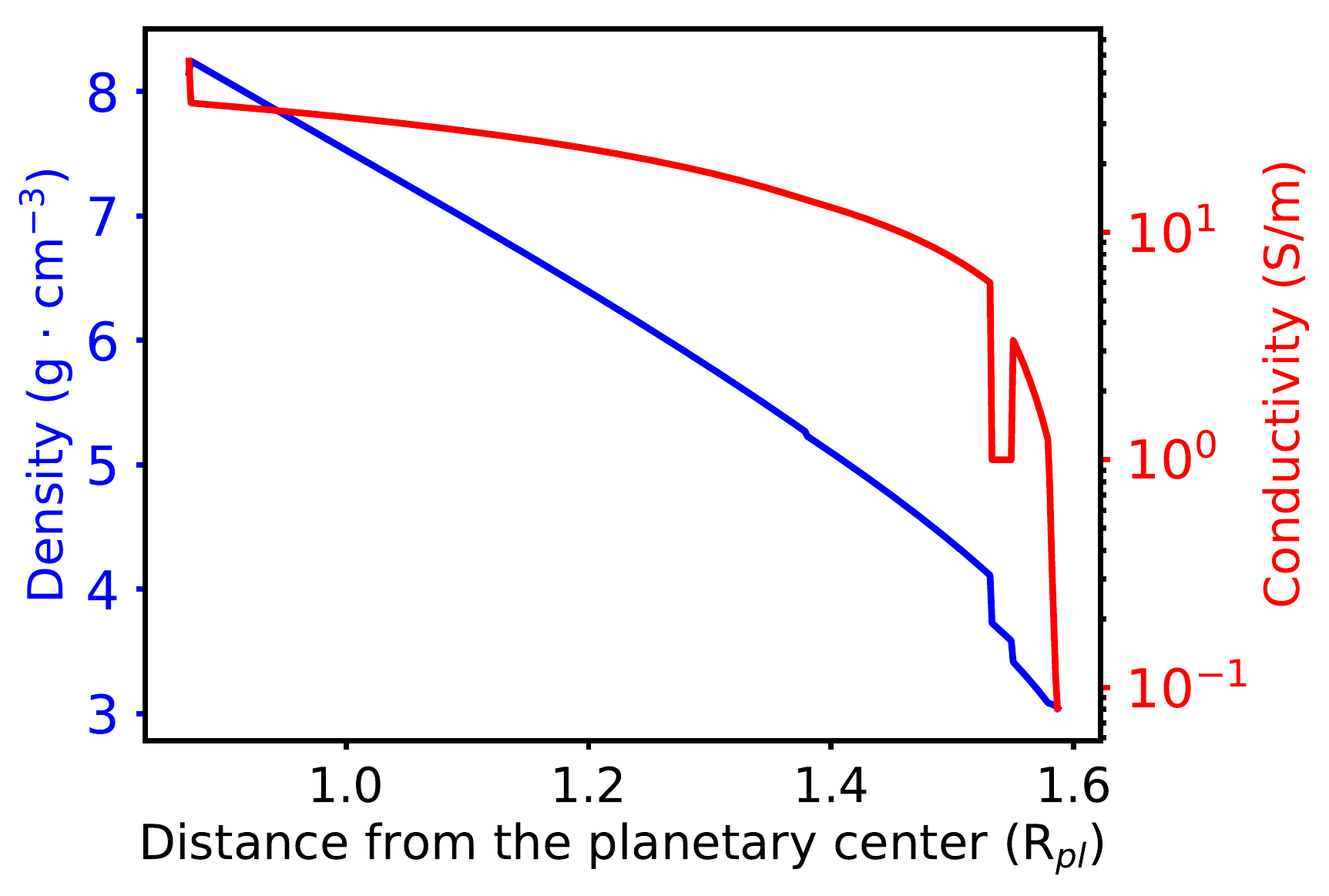}
 \caption{Density and conductivity profiles of the super-Earth HD~3167b, assuming a composition of 40\% iron and 60\% silicates and a surface temperature of 1759~K calculated with the code CHIC \citep{Noack16CHIC}. For comparison, the Earth has an iron content of about 32\%. The profile is adopted from \citet{GK20}. }
  \label{f_profiles}
\end{figure}

The conductivity profile of a planet's mantle is an important factor that influences the energy release by induction heating. Electrical conductivity and the frequency of variation of the magnetic field determine the skin depth, $\delta$, which is the penetration depth of the electromagnetic field into the medium, defined as the level where its amplitude decreases by a factor of $e$. In the approximation of a highly conductive medium ($4 \pi \sigma \gg \epsilon \omega$), which is applicable for the parameter range of interest, the skin depth is given by $\delta = c / \sqrt{2 \pi \sigma \mu \omega}$, where $c$ is the speed of light, $\sigma$ is the electrical conductivity of the medium, $\mu$ is the magnetic permeability, $\epsilon$ is the permittivity, and $\omega$ is the frequency of the field change. We assume $\mu = \epsilon = 1$ because the Curie temperature at which the materials lose their magnetic properties (about 600$^\circ$ for olivine) is reached at a depth of only 20--90~km in our model, and this depth depends on the assumed planet and mantle parameters (see discussion in \citealp{K17}).  The skin depth drastically decreases in the deeper layers, meaning that the varying magnetic field cannot penetrate the entire volume of the mantle. The conductivity is given by
\begin{equation}
    \sigma = \sigma_0 \exp\left( - \frac{\Delta H}{k_B T} \right),
    \label{eq_sigma}
\end{equation}
where $k_B$ is the Boltzmann constant and $T$ the local temperature. The values of $\sigma_0$ and $\Delta H$ for different mineral phases are taken from experimental studies \citep{Xu00,Yoshino08,Yoshino13}. Figure~\ref{f_profiles} shows density and conductivity profiles calculated for HD~3167b with the interior structure module in the code Coupling Habitability, Interior, and Crust (CHIC; \citealp{Noack16CHIC}). We assumed an Earth-like mantle mineralogy composed of Mg, Fe, Si, and O, which make up most of the Earth's mantle. Phase transitions (olivine$\rightarrow$wadsleyite, wadsleyite$\rightarrow$ringwoodite, ringwoodite$\rightarrow$perovskite mixed with magnesiowustite, and perovskite-post-perovskite) are clearly visible in the interior profile shown in Fig.~\ref{f_profiles}.

We calculated the amount of outgassing from the mantle of HD~3167b using the 2D mantle convection module in CHIC \citep{Noack16CHIC}. The code models the mantle convection and related magmatic events in a regional 2D spherical annulus geometry for a compressible mantle \citep{Noack17}. It also traces the amount of melt and its evolution in time. 

Here we present one example case to demonstrate the effect of the induction heating. The core was initially super-heated and cooled over time. We added Earth-like decaying radiogenic heat sources that  heat  the mantle. We assumed that the planet is in the stagnant lid regime. Plate tectonics would allow for strong outgassing; however, a stagnant lid is more likely for HD~3167b due to its high surface temperature \citep{Lenardic08}. We set the surface temperature in the convection simulations to 1000~K, which is lower than the equilibrium temperature of HD~3167b of 1759~K. A higher surface temperature cannot be modeled due to numerical constraints. Since HD~3167b is likely tidally locked, the temperature of 1759~K may be reached only in some regions of one hemisphere. Therefore, a surface temperature of 1000~K is a reasonable approximation for the global average temperature. Simulations start directly after the solidification of the magma ocean.

Furthermore, we assume that initially 200 wt-ppm of water are stored in the mantle, and consider a moderately oxidized mantle (IW+2) with initially 100 wt-ppm CO$_2$ being stored in the mantle in the form of graphite. The main outgassing products in our simulations are H$_2$, H$_2$O, CO, and CO$_2$ based on the local oxygen fugacity and volatile abundances in the melt \citep{OrtenziUReview}.

\section{Results}
\label{results}

\begin{figure}
\includegraphics[width=1.0\columnwidth]{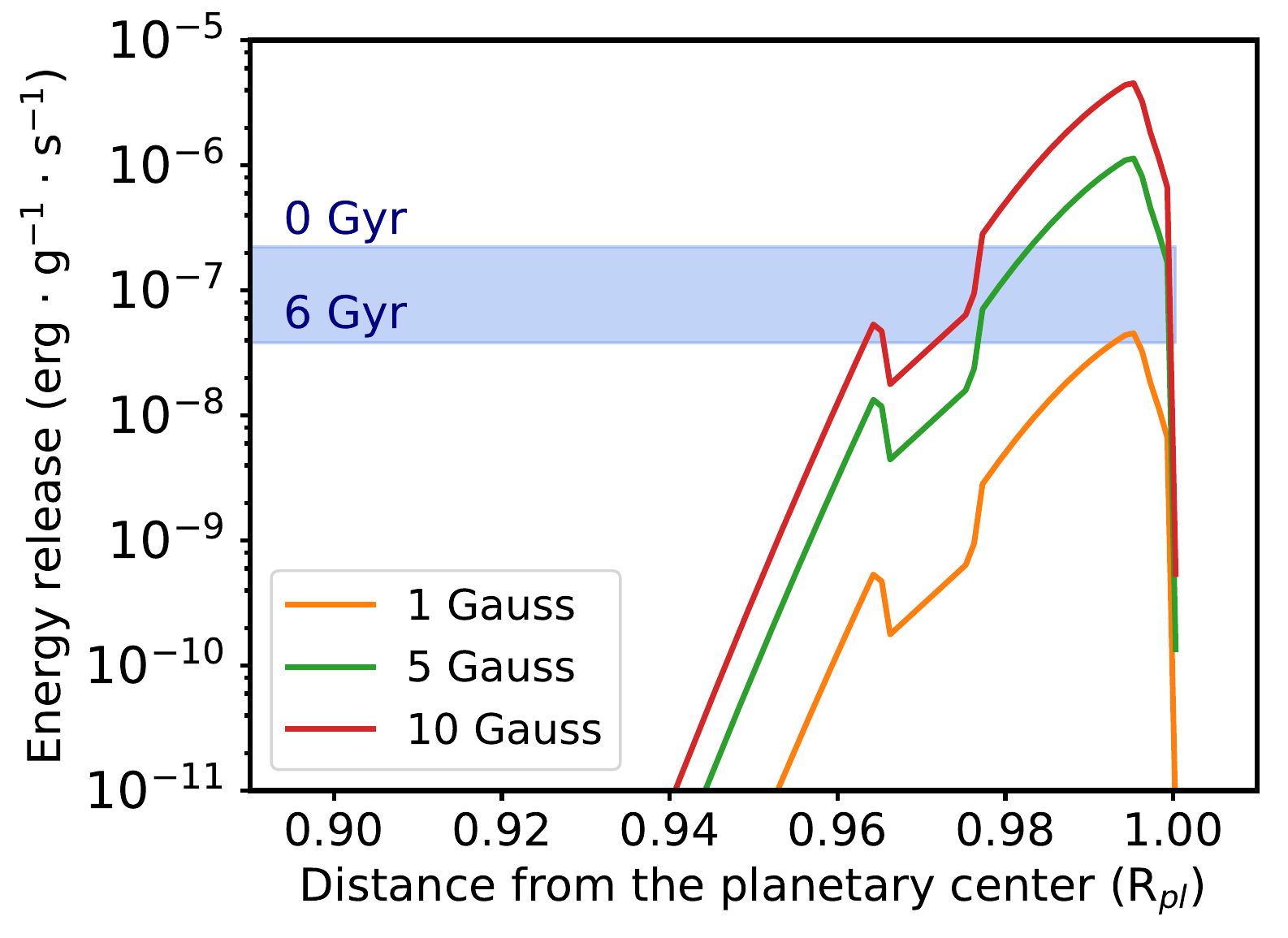}
 \caption{Energy release inside the planet HD~3167b assuming a conductivity profile with an iron mass fraction of 40\% and various strengths of the stellar magnetic field. The orbital inclination is 85$^\circ$. The energy is released in the upper 10\% of the planet's mantle. The shaded region shows the radiogenic heat sources at the formation time of the system (upper limit) and at the age of 6~Gyr (lower limit). }
  \label{f_heating}
\end{figure}

Figure~\ref{f_heating} shows the calculated induction heating inside the planetary mantle assuming different dipole strengths of the global magnetic field of the star HD~3167 of 1, 5, and 10 gauss. Young fast rotating Sun-like stars can generate strong global magnetic fields reaching around 100~G. As stars age their rotation decelerates and their magnetic fields become weaker \citep{Guedel07,Vidotto14}. Older Sun-like stars also exhibit cyclic behavior when the strengths of their magnetic fields and their dipole components vary periodically, with periods of years \citep{Fares13,Fares17,BoroSaikia18}. However, if a planet orbits very close to its host star, even an older star with a global dipole component of 3--10~G can generate a magnetic field of $\sim$0.1--0.5~G at the planet's orbit. A global dipole field of 10 gauss and less is typical for old and inactive stars such as HD~3167, which has a rotation period of 26~days \citep{Vidotto14}. For comparison, the global solar dipole field varies in the range of 1--7~G. The magnetic field at the planet's orbit was calculated as in \citet{K17} and was equal to 0.05, 0.26, and 0.52~G for the star's global dipole field of 1, 5, and 10 gauss, respectively. The total energy release due to induction heating in these three cases was equal to $1.8 \times 10^{19}$, $4.5 \times 10^{20}$, and $1.8 \times 10^{21}$~erg~s$^{-1}$, respectively. As one can see, the energy release scales with the magnetic field to the power of two.

\begin{figure*}
\includegraphics[width=0.9\columnwidth]{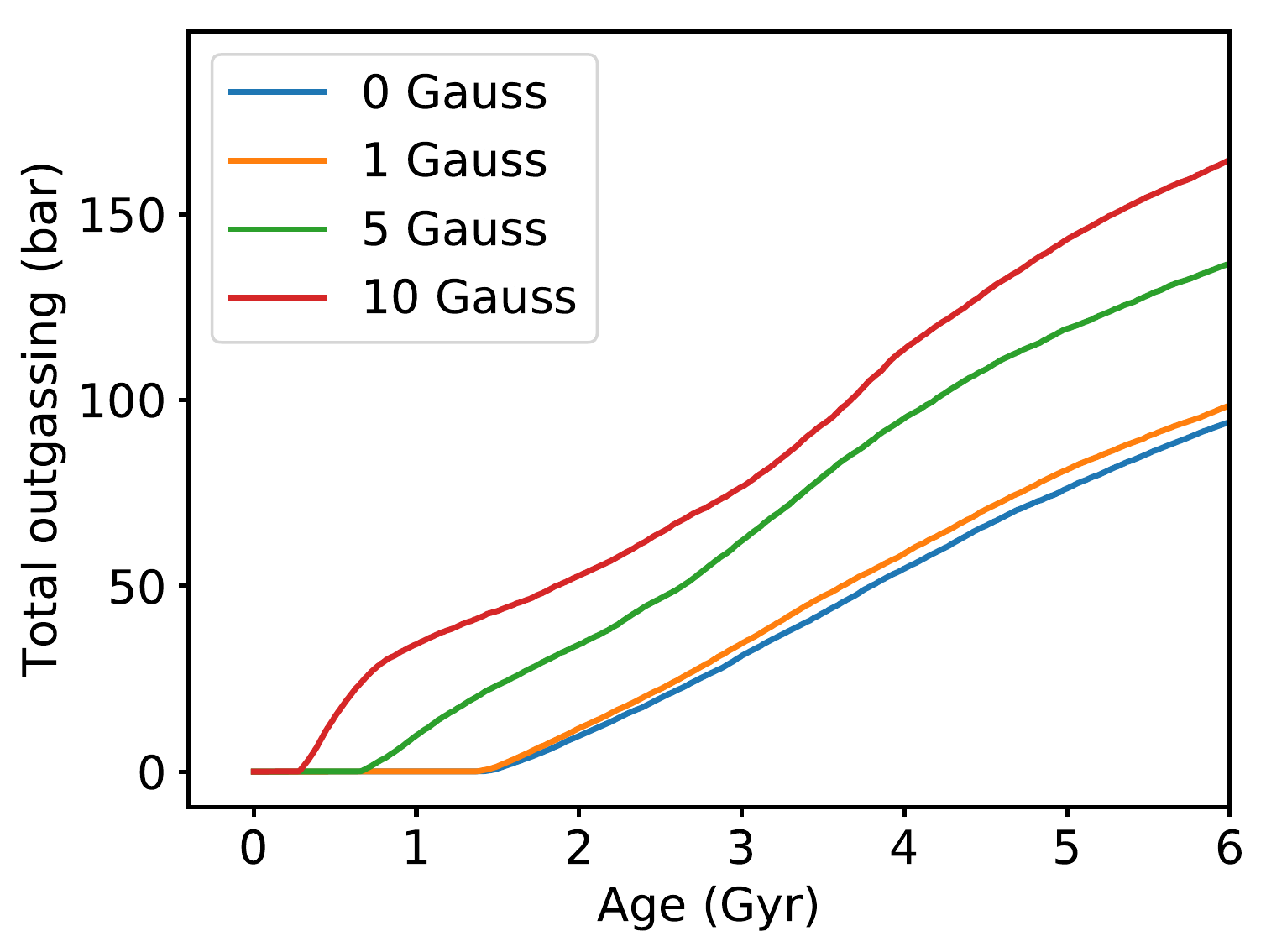}
\includegraphics[width=0.9\columnwidth]{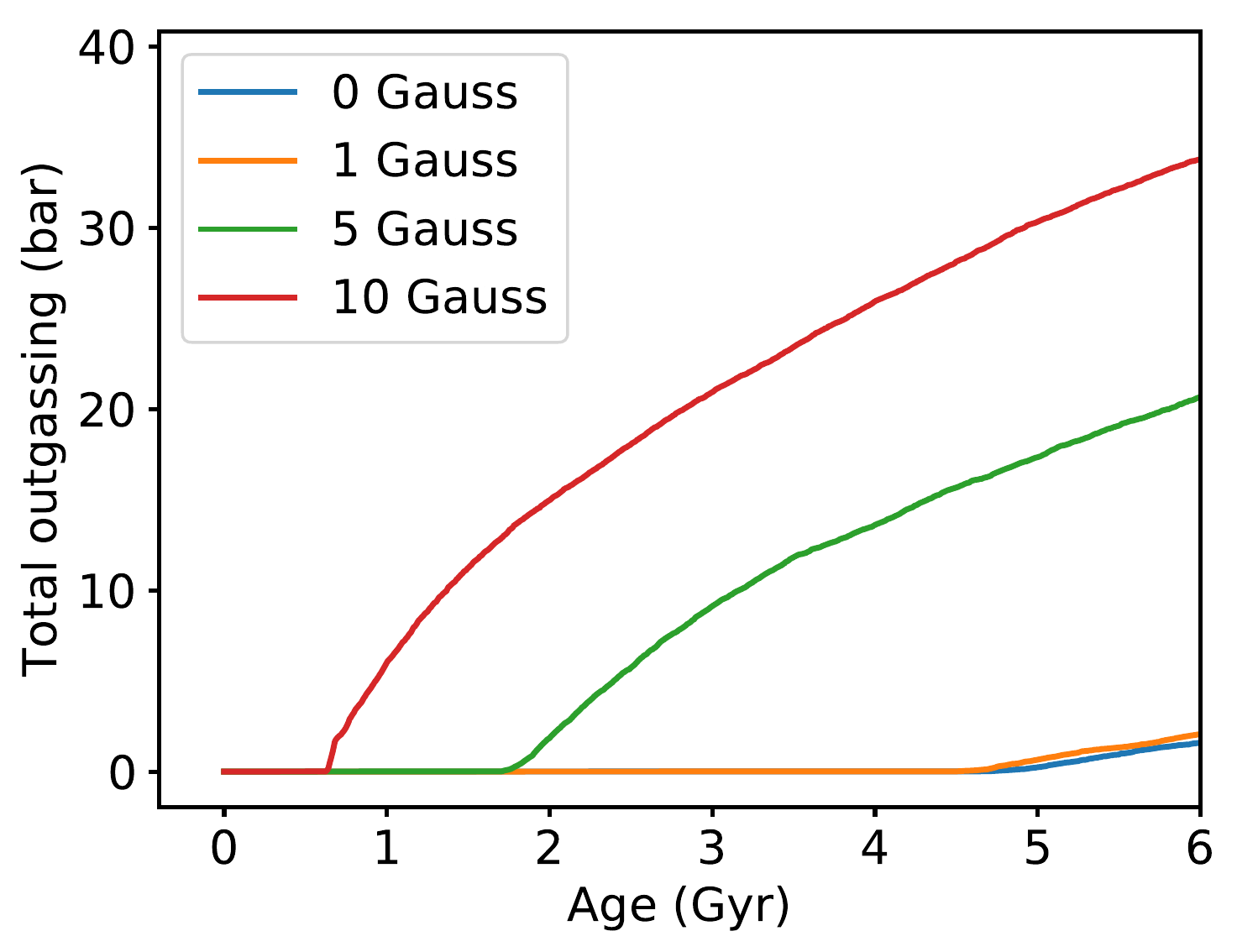}
 \caption{Total outgassing of CO$_2$, CO, H$_2$O, and H$_2$ from HD~3167b assuming the heating shown in Fig.~\ref{f_heating} for the magnetic field of the host star of 0, 1, 5, and 10~G. Induction heating leads to a much earlier onset of volcanism on the planet and increases the outgassing by several tens of bar. Left panel: Low (Earth-like) viscosity case. Right panel: Mantle viscosity is increased by a factor of 100.}
  \label{f_outgassing}
\end{figure*}

Figure~\ref{f_outgassing} shows the evolution of the total amount of volcanic outgassing for a global dipole field of HD~3167 of 0 (no induction heating), 1, 5, and 10 gauss, assuming Earth-like mantle viscosity (left panel) and viscosity enhanced by a factor of 100 (following \citealp{Tackley13}; right panel). Even a moderate global stellar magnetic field of 5 gauss leads to a much earlier onset of the volcanic activity on the planet and significantly increases its magnitude in comparison to the case without induction heating (0 gauss). In the 0 gauss case,  heat is still produced in the mantle due to radioactive decay. In the case with increased viscosity without induction heating, it takes more than five gigayears for the outgassed atmosphere to accumulate a pressure of several bars. In that case, due to a thick lithosphere that forms in the absence of plate tectonics, only a small amount of melt can form and rise to the surface due to the high pressure at the bottom of the lithosphere \citep{Dorn18}. For an Earth-like viscosity, outgassing is possible due to high surface temperature and efficient mantle convection. However, induction heating increases the outgassing by several dozen bar, if the global stellar magnetic field exceeds 5 gauss. 

All cases that include induction heating show an earlier onset of volcanic activity and predict the formation of an atmosphere with a pressure of several tens of bars. Even a weak stellar magnetic field of 1 gauss leads to a minor increase in outgassing at later evolutionary stages, when the radiogentic heat sources decline. Considering the close proximity of HD~3167b to its host star, the atmosphere of this planet is likely thinner than the estimates shown in Fig.~\ref{f_outgassing} because these results do not take into account escape to space or any other atmospheric sinks. Global magnetic fields of 5 to 10 gauss are quite typical for stars with activity levels and ages similar to HD~3167 \citep{Vidotto14}. In summary, our results predict active volcanism and the presence of an atmosphere or at least an extended exosphere on HD~3167b in contrast to earlier studies that neglected induction heating \citep{Noack17,Dorn18}.

Exoplanets on inclined orbits are subject to tidal heating. However, the local energy release by tidal dissipation depends on the rheology and shear modulus the mantle \citep{Roberts08}. More energy is released in the lower mantle where the temperatures are higher than in the upper mantle. Therefore, tidal heating is unlikely to melt the mantle of a planet as massive as HD~3167b. Unlike tidal heating, induction heating preferentially heats the very upper part of the mantle. Our results confirm that in this case some melt is produced in the upper mantle, resulting in volcanic activity. From our simulations we see that induction heating is indeed a possible driver of volcanism on super-Earths, and that it is very efficient for close-in planets orbiting Sun-like stars. Another important contributing factor is likely  the high surface temperature, which makes the lithosphere thinner, which is why even in the case without induction heating some outgassing can occur in the stagnant lid regime, in contrast to the predictions by \citet{Noack17} and \citet{Dorn18}.

\section{Conclusions}
\label{conclusions}

Our simulations confirm the suggestion by \citet{GK20} that induction heating can enhance volcanic activity in massive super-Earths. These findings have very important implications for  observational missions such as ARIEL, JWST, and the ground-based facilities if future spectroscopical observations of these massive exoplanets reveal the presence of atmospheres or exospheres with compositions typical of volcanic outgassing. Previously, some signs of volcanic activity were detected in 55~Cnc~e \citep{Demory16,Ridden16}. We conclude that induction heating coupled with the high surface temperature can be an important driver of volcanic activity on rocky super-Earths with masses exceeding approximately four Earth masses. For very close-in planets with surface temperatures exceeding 1000 K, a thin mineral atmosphere can also form \citep{Ito15}. One can discriminate between the two atmosphere types by investigating the atmosphere's composition, with volcanically outgassed atmospheres dominated by CO$_2$, CO, N$_2$, H$_2$, H$_2$O, H$_2$S, S$_2$ \citep{Gaillard14}, and mineral atmospheres dominated  by Na, Mg, SiO, O, among others \citep{Miguel11,Ito15}.

In this article we did not take into account the interaction of the volcanic atmosphere with the ambient stellar wind plasma. Such interactions perturb the magnetic field in the vicinity of a planet and can alter the induction response generated in the mantle. Important effects include plasma mass loading from newly picked-up atmospheric ions, possible Alf\'en wave currents that can flow through the planet's ionosphere, diamagnetism from newly picked-up plasma (e.g., \citealp{Khurana98,Saur04}). Plasma interactions produce a more complicated field structure in the planet's vicinity; for instance, the field can become stronger upstream and weaker downstream. Induction effects take place and are observable even in bodies surrounded by plasma, as is the case for the Galilean satellites embedded into the flow of Jupiter's magnetospheric plasma \citep{Zimmer00,Khurana11,Roth17} and for Mars, which interacts with the solar wind flow. Recently, it has also been shown that electromagnetic induction can power a large water current in Europa's ocean \citep{Gissinger19}. On Mars, the induction response to variations in the interplanetary magnetic field has been measured, and has allowed us to reconstruct the conductivity profile of the planet's mantle \citep{Civet14}. On HD~3167b, inclusion of the plasma effects would likely lead to generation of a more complicated induced magnetic field inside the planet in comparison to the dipolar field considered here. A thorough investigation of these effects using MHD simulations and a wind model  similar to those existing in the literature (e.g., \citealp{Johnstone15,Cohen17,BoroSaikia20}) is necessary to shed light on the role played by plasma effects.

Recently, \citet{GK20} have observed HD3167b in-transit and out-of-transit  with UVES at the VLT in order to search for the presence of lines originating from the planet's exosphere with the aim of detecting a possible volcanic or mineral atmosphere. They have searched for lines such as the $\rm Na\,D_{1,2}$ and Ca{\small II}\,H\&K lines as well as numerous [S II], [S III], and [O III] lines, which  are tracers of volcanic activity. They derived upper limits of the ratios of the line flux to the stellar flux. The derived upper limits were $1.5\cdot10^{-3}$ for the Ca{\small II}\,H\&K lines and $7.2\cdot10^{-4}$ and $3.3\cdot10^{-4}$ and for the $\rm NaD_{1,2}$ lines, respectively.  The fact that these upper limits are approximately equal to previous detections in 55\,Cnc\,e by \citet{Ridden16} shows that not all super-Earths show these lines continuously and that they might be variable. This indicates that further observations,   at different wavelengths and with different instruments such as ARIEL and JWST, are necessary.

\begin{acknowledgements}

KK acknowledges the support by Austrian Science Fund (FWF) NFN project S116-N16 and   subproject S11604-N16. The authors thank the Erwin Schr\"odinger Institute (ESI) of the University of Vienna for hosting the meetings of the Thematic Program ``Astrophysical Origins: Pathways from Star Formation to Habitable Planets'' and Europlanet for providing additional support for this program.

\end{acknowledgements}

\bibliographystyle{aa} 
\bibliography{HD3167b_references}

\begin{thebibliography}{42}
\expandafter\ifx\csname natexlab\endcsname\relax\def\natexlab#1{#1}\fi

\bibitem[{{Boro Saikia} {et~al.}(2020){Boro Saikia}, {Jin}, {Johnstone},
  {L{\"u}ftinger}, {G{\"u}del}, {Airapetian}, {Kislyakova}, \&
  {Folsom}}]{BoroSaikia20}
{Boro Saikia}, S., {Jin}, M., {Johnstone}, C.~P., {et~al.} 2020, \aap, 635,
  A178

\bibitem[{{Boro Saikia} {et~al.}(2018){Boro Saikia}, {Lueftinger}, {Jeffers},
  {Folsom}, {See}, {Petit}, {Marsden}, {Vidotto}, {Morin}, {Reiners}, {Guedel},
  \& {BCool Collaboration}}]{BoroSaikia18}
{Boro Saikia}, S., {Lueftinger}, T., {Jeffers}, S.~V., {et~al.} 2018, \aap,
  620, L11

\bibitem[{{Civet} \& {Tarits}(2014)}]{Civet14}
{Civet}, F. \& {Tarits}, P. 2014, Earth, Planets, and Space, 66, 85

\bibitem[{{Cohen}(2017)}]{Cohen17}
{Cohen}, O. 2017, \apj, 835, 220

\bibitem[{{Dalal} {et~al.}(2019){Dalal}, {H{\'e}brard}, {Lecavelier des
  {\'E}tangs}, {Petit}, {Bourrier}, {Laskar}, {K{\"o}nig}, \&
  {Correia}}]{Dalal19}
{Dalal}, S., {H{\'e}brard}, G., {Lecavelier des {\'E}tangs}, A., {et~al.} 2019,
  \aap, 631, A28

\bibitem[{{Demory} {et~al.}(2016){Demory}, {Gillon}, {Madhusudhan}, \&
  {Queloz}}]{Demory16}
{Demory}, B.-O., {Gillon}, M., {Madhusudhan}, N., \& {Queloz}, D. 2016, \mnras,
  455, 2018

\bibitem[{{Dorn} {et~al.}(2018){Dorn}, {Noack}, \& {Rozel}}]{Dorn18}
{Dorn}, C., {Noack}, L., \& {Rozel}, A.~B. 2018, \aap, 614, A18

\bibitem[{{Edwards} {et~al.}(2019){Edwards}, {Mugnai}, {Tinetti}, {Pascale}, \&
  {Sarkar}}]{Edwards19}
{Edwards}, B., {Mugnai}, L., {Tinetti}, G., {Pascale}, E., \& {Sarkar}, S.
  2019, \aj, 157, 242

\bibitem[{{Fares} {et~al.}(2017){Fares}, {Bourrier}, {Vidotto}, {Moutou},
  {Jardine}, {Zarka}, {Helling}, {Lecavelier des Etangs}, {Llama}, {Louden},
  {Wheatley}, \& {Ehrenreich}}]{Fares17}
{Fares}, R., {Bourrier}, V., {Vidotto}, A.~A., {et~al.} 2017, \mnras, 471, 1246

\bibitem[{{Fares} {et~al.}(2013){Fares}, {Moutou}, {Donati}, {Catala},
  {Shkolnik}, {Jardine}, {Cameron}, \& {Deleuil}}]{Fares13}
{Fares}, R., {Moutou}, C., {Donati}, J.-F., {et~al.} 2013, \mnras, 435, 1451

\bibitem[{{Gaillard} \& {Scaillet}(2014)}]{Gaillard14}
{Gaillard}, F. \& {Scaillet}, B. 2014, Earth and Planetary Science Letters,
  403, 307

\bibitem[{{Gandolfi} {et~al.}(2017){Gandolfi}, {Barrag{\'a}n}, {Hatzes},
  {Fridlund}, {Fossati}, {Donati}, {Johnson}, {Nowak}, {Prieto-Arranz},
  {Albrecht}, {Dai}, {Deeg}, {Endl}, {Grziwa}, {Hjorth}, {Korth}, {Nespral},
  {Saario}, {Smith}, {Antoniciello}, {Alarcon}, {Bedell}, {Blay}, {Brems},
  {Cabrera}, {Csizmadia}, {Cusano}, {Cochran}, {Eigm{\"u}ller}, {Erikson},
  {Gonz{\'a}lez Hern{\'a}ndez}, {Guenther}, {Hirano}, {Su{\'a}rez
  Mascare{\~n}o}, {Narita}, {Palle}, {Parviainen}, {P{\"a}tzold}, {Persson},
  {Rauer}, {Saviane}, {Schmidtobreick}, {Van Eylen}, {Winn}, \&
  {Zakhozhay}}]{gandolfi17}
{Gandolfi}, D., {Barrag{\'a}n}, O., {Hatzes}, A.~P., {et~al.} 2017, \aj, 154,
  123

\bibitem[{{Gissinger} \& {Petitdemange}(2019)}]{Gissinger19}
{Gissinger}, C. \& {Petitdemange}, L. 2019, Nature Astronomy, 3, 401

\bibitem[{{Godolt} {et~al.}(2019){Godolt}, {Tosi}, {Stracke}, {Grenfell},
  {Ruedas}, {Spohn}, \& {Rauer}}]{Godolt19}
{Godolt}, M., {Tosi}, N., {Stracke}, B., {et~al.} 2019, \aap, 625, A12

\bibitem[{{G{\"u}del}(2007)}]{Guedel07}
{G{\"u}del}, M. 2007, Living Reviews in Solar Physics, 4, 3

\bibitem[{{G\"unther} \& {Kislyakova}(2020)}]{GK20}
{G\"unther}, E.~W. \& {Kislyakova}, K.~G. 2020, \mnras, 491, 3974

\bibitem[{{Ito} {et~al.}(2015){Ito}, {Ikoma}, {Kawahara}, {Nagahara},
  {Kawashima}, \& {Nakamoto}}]{Ito15}
{Ito}, Y., {Ikoma}, M., {Kawahara}, H., {et~al.} 2015, \apj, 801, 144

\bibitem[{{Johnstone}(2012)}]{JohnstoneThesis}
{Johnstone}, C.~P. 2012, PhD thesis, University of St Andrews, email:
  colin.johnstone@univie.ac.at

\bibitem[{{Johnstone} {et~al.}(2015){Johnstone}, {G{\"u}del}, {L{\"u}ftinger},
  {Toth}, \& {Brott}}]{Johnstone15}
{Johnstone}, C.~P., {G{\"u}del}, M., {L{\"u}ftinger}, T., {Toth}, G., \&
  {Brott}, I. 2015, \aap, 577, A27

\bibitem[{{Khurana} {et~al.}(2011){Khurana}, {Jia}, {Kivelson}, {Nimmo},
  {Schubert}, \& {Russell}}]{Khurana11}
{Khurana}, K.~K., {Jia}, X., {Kivelson}, M.~G., {et~al.} 2011, Science, 332,
  1186

\bibitem[{{Khurana} {et~al.}(1998){Khurana}, {Kivelson}, {Stevenson},
  {Schubert}, {Russell}, {Walker}, \& {Polanskey}}]{Khurana98}
{Khurana}, K.~K., {Kivelson}, M.~G., {Stevenson}, D.~J., {et~al.} 1998, \nat,
  395, 777

\bibitem[{{Kislyakova} {et~al.}(2018){Kislyakova}, {Fossati}, {Johnstone},
  {Noack}, {L{\"u}ftinger}, {Zaitsev}, \& {Lammer}}]{K18}
{Kislyakova}, K.~G., {Fossati}, L., {Johnstone}, C.~P., {et~al.} 2018, \apj,
  858, 105

\bibitem[{{Kislyakova} {et~al.}(2017){Kislyakova}, {Noack}, {Johnstone},
  {Zaitsev}, {Fossati}, {Lammer}, {Khodachenko}, {Odert}, \& {Guedel}}]{K17}
{Kislyakova}, K.~G., {Noack}, L., {Johnstone}, C.~P., {et~al.} 2017, Nature
  Astronomy, 1, 878

\bibitem[{{Lang} {et~al.}(2012){Lang}, {Jardine}, {Donati}, {Morin}, \&
  {Vidotto}}]{Lang12}
{Lang}, P., {Jardine}, M., {Donati}, J.-F., {Morin}, J., \& {Vidotto}, A. 2012,
  \mnras, 424, 1077

\bibitem[{Lenardic {et~al.}(2008)Lenardic, Jellinek, \& Moresi}]{Lenardic08}
Lenardic, A., Jellinek, A., \& Moresi, L.-N. 2008, Earth and Planetary Science
  Letters, 271, 34

\bibitem[{{McEwen} {et~al.}(1998){McEwen}, {Keszthelyi}, {Spencer}, {Schubert},
  {Matson}, {Lopes-Gautier}, {Klaasen}, {Johnson}, {Head}, {Geissler},
  {Fagents}, {Davies}, {Carr}, {Breneman}, \& {Belton}}]{McEwen98}
{McEwen}, A.~S., {Keszthelyi}, L., {Spencer}, J.~R., {et~al.} 1998, Science,
  281, 87

\bibitem[{{Miguel} {et~al.}(2011){Miguel}, {Kaltenegger}, {Fegley}, \&
  {Schaefer}}]{Miguel11}
{Miguel}, Y., {Kaltenegger}, L., {Fegley}, B., \& {Schaefer}, L. 2011, \apjl,
  742, L19

\bibitem[{{Morin} {et~al.}(2010){Morin}, {Donati}, {Petit}, {Delfosse},
  {Forveille}, \& {Jardine}}]{Morin10}
{Morin}, J., {Donati}, J.-F., {Petit}, P., {et~al.} 2010, \mnras, 407, 2269

\bibitem[{Noack {et~al.}(2016)Noack, Rivoldini, \& Van~Hoolst}]{Noack16CHIC}
Noack, L., Rivoldini, A., \& Van~Hoolst, T. 2016, International Journal On
  Advances in Systems and Measurements, 9, 66

\bibitem[{{Noack} {et~al.}(2017){Noack}, {Rivoldini}, \& {Van
  Hoolst}}]{Noack17}
{Noack}, L., {Rivoldini}, A., \& {Van Hoolst}, T. 2017, Physics of the Earth
  and Planetary Interiors, 269, 40

\bibitem[{{Ortenzi} {et~al.}(2019){Ortenzi}, {Noack}, {Sohl}, {Guimond},
  {Grenfell}, {Dorn}, {Schmidt}, {Vulpius}, {Katyal}, {Kitzmann}, \&
  {Rauer}}]{OrtenziUReview}
{Ortenzi}, G., {Noack}, L., {Sohl}, F., {et~al.} 2019, Nature Scientific
  Reports, submitted

\bibitem[{{Parkinson}(1983)}]{Parkinson83}
{Parkinson}, W.~D. 1983, {Introduction to Geomagnetism} (Scottish Academic
  Press Ltd (May 1983), ISBN-10: 0707302927)

\bibitem[{{Ridden-Harper} {et~al.}(2016){Ridden-Harper}, {Snellen}, {Keller},
  {de Kok}, {Di Gloria}, {Hoeijmakers}, {Brogi}, {Fridlund}, {Vermeersen}, \&
  {van Westrenen}}]{Ridden16}
{Ridden-Harper}, A.~R., {Snellen}, I.~A.~G., {Keller}, C.~U., {et~al.} 2016,
  \aap, 593, A129

\bibitem[{{Roberts} \& {Nimmo}(2008)}]{Roberts08}
{Roberts}, J.~H. \& {Nimmo}, F. 2008, \icarus, 194, 675

\bibitem[{{Roth} {et~al.}(2017){Roth}, {Saur}, {Retherford}, {Bl{\"o}cker},
  {Strobel}, \& {Feldman}}]{Roth17}
{Roth}, L., {Saur}, J., {Retherford}, K.~D., {et~al.} 2017, Journal of
  Geophysical Research (Space Physics), 122, 1903

\bibitem[{{Saur} {et~al.}(2004){Saur}, {Neubauer}, {Connerney}, {Zarka}, \&
  {Kivelson}}]{Saur04}
{Saur}, J., {Neubauer}, F.~M., {Connerney}, J.~E.~P., {Zarka}, P., \&
  {Kivelson}, M.~G. 2004, {Plasma interaction of Io with its plasma torus}, ed.
  F.~{Bagenal}, T.~E. {Dowling}, \& W.~B. {McKinnon}, Vol.~1, 537--560

\bibitem[{{Tackley} {et~al.}(2013){Tackley}, {Ammann}, {Brodholt}, {Dobson}, \&
  {Valencia}}]{Tackley13}
{Tackley}, P.~J., {Ammann}, M., {Brodholt}, J.~P., {Dobson}, D.~P., \&
  {Valencia}, D. 2013, \icarus, 225, 50

\bibitem[{{Vidotto} {et~al.}(2014){Vidotto}, {Gregory}, {Jardine}, {Donati},
  {Petit}, {Morin}, {Folsom}, {Bouvier}, {Cameron}, {Hussain}, {Marsden},
  {Waite}, {Fares}, {Jeffers}, \& {do Nascimento}}]{Vidotto14}
{Vidotto}, A.~A., {Gregory}, S.~G., {Jardine}, M., {et~al.} 2014, \mnras, 441,
  2361

\bibitem[{{Xu} {et~al.}(2000){Xu}, {Shankland}, \& {Poe}}]{Xu00}
{Xu}, Y., {Shankland}, T.~J., \& {Poe}, B.~T. 2000, \jgr, 105, 27

\bibitem[{{Yoshino} \& {Katsura}(2013)}]{Yoshino13}
{Yoshino}, T. \& {Katsura}, T. 2013, Annual Review of Earth and Planetary
  Sciences, 41, 605

\bibitem[{{Yoshino} {et~al.}(2008){Yoshino}, {Manthilake}, {Matsuzaki}, \&
  {Katsura}}]{Yoshino08}
{Yoshino}, T., {Manthilake}, G., {Matsuzaki}, T., \& {Katsura}, T. 2008, \nat,
  451, 326

\bibitem[{{Zimmer} {et~al.}(2000){Zimmer}, {Khurana}, \& {Kivelson}}]{Zimmer00}
{Zimmer}, C., {Khurana}, K.~K., \& {Kivelson}, M.~G. 2000, \icarus, 147, 329

\end{thebibliography}

\end{document}